\documentstyle[prd,aps,epsfig,amssymb]{revtex}
\begin{document}
\draft 
\preprint{} 
\title{ The Spectrum of a Binding System for a Heavy Quark
        with an Anti-Sbottom or for a Sbottom
        and Anti-Sbottom Pair }

\author{Chao-Hsi Chang$^{a,b}$, Jian Ying Cui$^c$ and Jin Min Yang$^b$}

\address{ $^a$ CCAST (World Laboratory), P.O.Box 8730, Beijing 100080, China}
\address{ $^b$ Institute of Theoretical Physics, Academia Sinica, Beijing 100080, China} 
\address{ $^c$ Department of Physics, Yantai University, Yantai, Shandong 264005, China}

\maketitle

\begin{abstract}

Since long-lived light bottom squark (sbottom) and its anti-particle
with a mass close to the bottom quark have not been excluded by
experiments so far, we consider such a sbottom to combine
with its anti-particle to form a color singlet meson-like bound state or
to combine with a common anti-quark to form a fermion-like one, or
accordingly their anti-particles to form an anti-particle bound system.
Namely we calculate the low-lying spectrum of the systems
based on QCD inspired potential model. To be as relativistic
as possible, we start with the framework of Bethe-Salpeter (BS) equation
even for non-relativistic binding systems. Finally, we obtain the requested
spectrum by constructing general forms of the BS wave functions and solving
the BS equations under instantaneous approximation.
\end{abstract}


\section{ Introduction}

The supersymmetric minimum extended standard model, i.e. Minimum
Supersymetric Standard Model (MSSM), which have consequences at or
lower than weak-scale \cite{s1}, is arguably the most promising
candidate for physics beyond the Standard  Model (SM). Generally
this theory predicts existence of a super partner corresponding to
each particle of SM, and these `partners', the supersymmetric
particles (sparticles), should not be heavier than $O(1)$ TeV.
Thus the sparticles should be accessible at the exist and
constructing colliders such as Tevatron and LHC. In recent years,
great deal of effort has been made to search for SUSY.
Unfortunately, no direct signal for SUSY has been observed so
far and some lower mass bounds have been established for
sparticles. Based on the experimental results at LEP \cite{s2,s3}
and Tevatron \cite{s4,s5}, squarks must be heavier than about 100
GeV. However, most of the analysis of experimental searches for
the sparticles is performed with model-dependent assumptions and
rely on a large missing energy cut. Hence the quoted bounds may be
escaped if some of the assumptions are relaxed. In particuliar, a
long-lived light sbottom, $\tilde b$, with mass close to the bottom
quark, has not been excluded by experiments. Some analyses
~\cite{zbb} showed that if the light $\tilde{b}$ is an appropriate
admixture of left-handed and right-handed bottom squarks, its
coupling to $Z$ boson can be small enough to avoid LEP-I $Z$ decay
bounds. In addition, a scenario with light gluino and long-lived
light sbottom, $\tilde b$, with mass close to the bottom quark,
was proposed in \cite{berger}, with which the excess of measured
$b\bar{b}$ pair production in hadron collision over QCD
theoretical prediction by a factor two is explained
successfully \footnote{In this case, i.e., gluino and the sbottom
$\tilde b$ are both light, the loop effects of such a sbottom may 
cause significant effects in the Z-peak observables \cite{z-peak}.
Thus more care should be paid in globally fitting the parameters 
of the model.}. The CLEO exclusion of a $\tilde b$ with mass 3.5 to 4.5
GeV ~\cite{CLEO} can also be loosed even avoided, since their
analysis depends on the assumption for semi-leptonic decays of the
light sbottom. Moreover, since sbottom is a scalar, based on the
spin freedom counting only, its pair production rate at colliders
will be smaller than the bottom quark by a factor four, so the
sbottom samples must be rarer than those of bottom quark in
experiments. Therefore, a light sbottom and its anti-particle, in
fact, are not excluded.

In contrary, it is interesting to point out
that some experiments seemingly favor such a
light sbottom. The ALEPH collaboration has reported experimental
hints for a light sbottom with a mass around 4 GeV and lifetime of
$1$ ps \cite{s7}. A recent reanalysis of old anomaly in the MARK-I
data for cross section of  $e^+e^-\rightarrow~ hadrons$ shows that
the existence of such a light sbottom can bring the measured cross
section into agreement with the theoretical prediction \cite{s6}.

The phenomenology of a very light sbottom has been studied by many
authors recently \cite{s8}. If such a light sbottom indeed exist,
new meson-like bound states by a pair of the sbottoms
($\tilde{b}\bar{\tilde{b}}$) and fermion-like ones by the sbottom
plus an ordinary quark ($Q\bar{\tilde{b}}$)\footnote{Here we would
like to restrict ourselves to consider the systems
($Q\bar{\tilde{b}}$), an `ordinary' heavy quark to bind with the
light sbottom, because with the restriction the systems are surely
non-relativistic, thus the well-tested knowledge for heavy-heavy
quark systems $(Q\bar{Q})$ can be used as solid reference for the
present study. Here $Q(\bar{Q})$ denotes $b(\bar{b})$ or
$c(\bar{c})$ only.} may be formed. In this paper we study these
new binding systems in the framework of BS equation, because this
equation has been successfully used in the study of the
heavy-heavy $(Q\bar{Q})$ bound states under some
approximation \footnote{Note here that in Ref.\cite{s13}, a
similar binding system $(Q\bar{\tilde{t_1}})$ for a light
anti-top-squark and a heavy quark is considered, but it is in
Chinese and some slight different approximation is taken.}.

In Sec.\ref{sec2} we give the BS equations for the
binding systems ($Q\bar{\tilde{b}})$ and ($\tilde{b}\bar{\bar{b}}$). We
construct the BS kernel by calculating the lowest order Feynman diagrams
and by phenomenological considerations similar to the cases of
$(Q\bar{Q})$ systems, i.e. the QCD inspired kernel. Then we start with
the general formulations of the BS wave functions for these bound states
to establish the coupling equations for the components of the BS wave
functions accordingly. In Sec. \ref{sec3} we choose the parameters in
our calculation by fitting the spectrum of $c\bar{c}$ and $b\bar{b}$ bound
states. Then we numerically solve the the BS equations for the systems
($Q\bar{\tilde{b}}$) and ($\tilde{b}\bar{\bar{b}}$). Finally
we present the numerical results and some discussions.

\section{The relevant Bethe-Salpeter (BS) equations}
\label{sec2}

\subsection{The bound state of a quark and a sbottom}
Let $|P\rangle$ (momentum $P_\mu$ and mass $M: P^2=M^2$)
denotes a bound state of a heavy quark $Q$ with
a light anti-sbottom ($\bar{\tilde{b}}$), $Q(x)$ denotes the heavy quark
field and $\bar{\tilde{b}}(x)$ the light anti-sbottom field,
where the color indices are suppressed. Then
the Bethe-Salpeter (BS) wave function of this bound state is defined as

\begin{equation}
\chi(P,x_1,x_2)=\langle 0 | T(Q(x_1)\bar{\tilde{b}}(x_2))| P
\rangle,
\end{equation}
The translation invariance of the system implies

\begin{equation}
\chi(P,x_1,x_2)=e^{iP\cdot X} \chi(P,x),
\end{equation}
where the center-of-mass and relative variables, and $X$ and $x$ are
coordinate of center mass system and the relative one of the two
components respectively:

\begin{equation}
X=\lambda_1 x_1+\lambda_2  x_2,~~~~~~~~ x=x_1-x_2.
\end{equation}
with $\displaystyle \lambda_i=\frac{m_i}{m_1+m_2} (i=1,2)$
and $m_1$, $m_2$ are the masses and momentums of the quark
and sbottom respectively.

We further
have the BS wave function in momentum space as

\begin{equation}
\chi(P,q)=\int \frac{d^4 x}{(2 \pi )^4}e^{-iqx}\chi(P,x),
\end{equation}
where $q$ is the relative momentum of the two components of the
bound state. Then in momentum space the BS equation is written as

\begin{equation}
\chi(P,q)=\frac{1}{{p_1}\!\!\!\! /  - m_1+i\epsilon}~
                        \frac{1}{p^2_2-m_2^2+i\epsilon}
          \int\frac{d^4 k}{(2\pi)^4}G(P,q,k)\chi(P,k),
\end{equation}
where $G(P,q,k)$ is the BS kernel which is defined as the sum of
all two-particle irreducible graphs. $m_1$, $p_1$ and $m_2$, $p_2$
are the masses and momentums of the quark and sbottom
respectively, so we have

\begin{equation}
\label{e2a}
 P=p_1+p_2,\hskip0.4in q=\lambda_2 p_1-\lambda_1 p_2.
\end{equation}
To fix the BS equation, we should decide the specific form of
the BS kernel. In the present case, the kernel is chosen based on
QCD spirit as a combination of a short distance part and a long
distance part:
\begin {equation}
G(P,q,k)=iG_s(P,q,k)+iG_l(P,q,k).
\end{equation}
The short distance part of the kernel, $G_s(P,q,k)$,
reasonably is of one-gluon-exchange, corresponding to
the Feynman diagram shown in Fig.~1.(a). $G_s$ is
precisely written as:
\begin {equation}
G_s(P,q,k)= \frac{4}{3}(4\pi \alpha_s)
    \frac{p\!\!\!/ _2 + p\!\!\!/ _4}{(p_2-p_4)^2 - \alpha^2},
\end{equation}
where $\frac{4}{3}$ is the color factor for color singlet bound state,
and $\alpha$ in the denominator is a small constant, which is introduced
here to avoid the infrared divergence in numerical calculation, and can be
considered as a fictitious gluon
mass. Based on QCD, here the strong coupling constant $\alpha_s$, as
heavy quarkonium, is chosen as
\begin{equation}
\alpha_s=\frac{12\pi}{27}\frac{1}{ln(a+\frac{(q-k)^2}{\Lambda^2_{QCD}})}.
\end{equation}
\begin{figure}
\vspace*{-4cm}
\epsfig{file=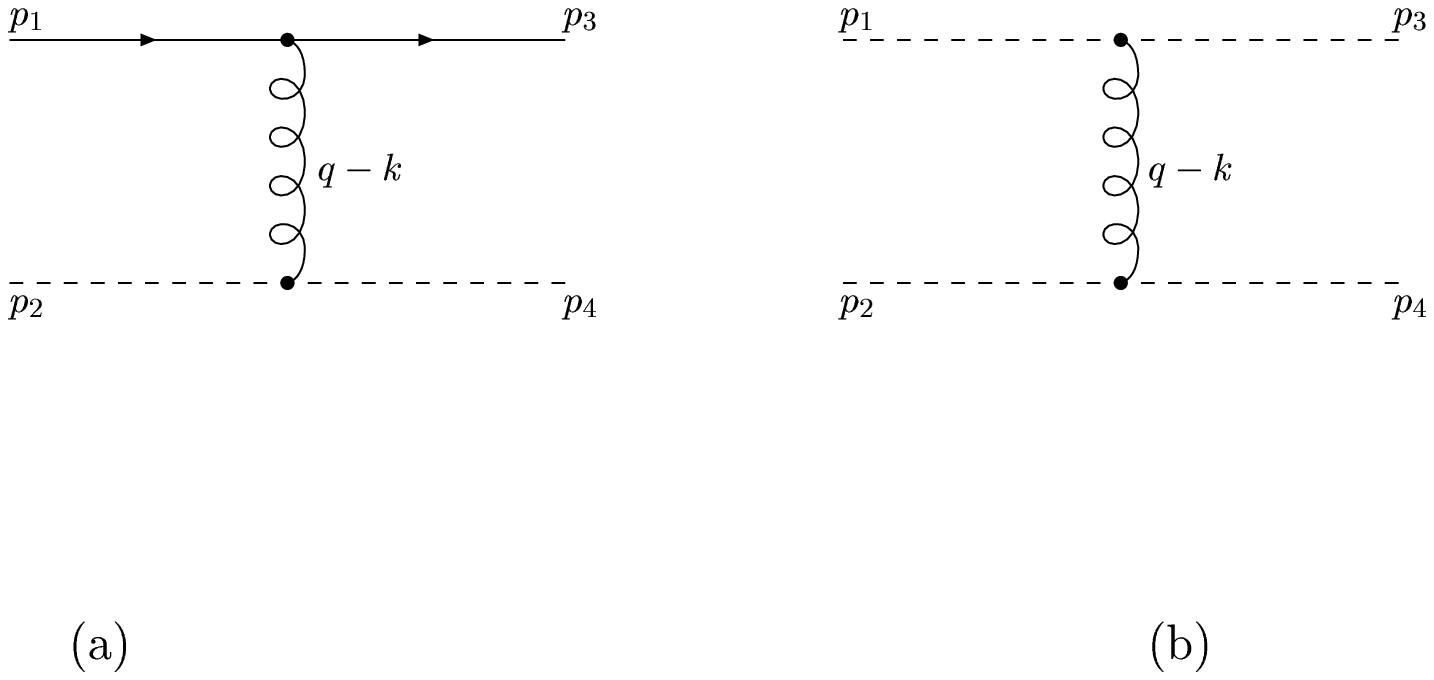,width=500pt,height=500pt}
\vspace*{-8cm}
\caption{The BS kernel from one-gloun-exchange: (a) for the system
$(b\bar{\tilde{b}})$; (b) for the system $(\tilde{b}\bar{\tilde{b}})$}.
\label{fig.1}
\end{figure}
Besides Eq.(6), because of momentum conservation we also have
\begin{equation}
P=p_3+p_4, \hskip0.4in
k=\lambda_2 p_3 - \lambda_1 p_4.
\end{equation}
Eq.(5) is fully Lorentz invariant, but it is hard to solve even
numerically. In actual investigation, as heavy quarkonium, an effective
method, the so-called instantaneous approximation, is made to
reduce the full equation to a three dimensional one. For this
purpose, the zero components of the momentums $p_i^0 (i=1,2,3,4)$
in the kernel should be fixed. Since we know that we are considering
weak binding systems, so the zero components are chosen, as an approximation,
as the on-shell values accordingly:

\begin{equation}
p^0_i=\sqrt{m_i^2+{\bf p}_i^2}\;, \;\;\;\; (i=1,2,3,4)\; .
\end {equation}
Then $G_s$ is rewritten as

\begin{equation}
G_s=G_s^{(1)}{\bf \gamma\cdot}({\bf q}+{\bf k})+G_s^{(2)}\gamma^0
\end{equation}
with

\begin{eqnarray}
G_s^{(1)}&=&\frac{4}{3}\frac{4\pi\alpha_s}
     {(E_2-E_2^\prime)^2-({\bf q}-{\bf k})^2+\alpha^2},\\
G_s^{(2)}&=&G_s^{(1)}(E_2+E_2^\prime),
\end{eqnarray}
where $E_i=\sqrt{m_i^2+{\bf q}^2}$ and
$E_i^\prime=\sqrt{m_i^2+{\bf k}^2}$ ($i=1,2$)\footnote{Since now on, we change
the notations a little that $p^0_i=E_i (i=1,2)$ and $p^0_{3,4}=E_{1,2}^\prime$.}.
As for the long distance part of the kernel, we have to construct it
phenomenologically. According to the experience of heavy quarkonium, ($Q\bar{Q}$)
bound state, we assume that it has Lorentz scalar property, and choose is as

\begin{equation}
G_l(P,q,k)= (2\pi)^3(2m_2)\{\frac{\lambda}{\alpha}\delta^3({\bf
q}-{\bf k})
          -\frac{\lambda}{\pi^2}\frac{1}{(({\bf q}-{\bf k})^2+\alpha^2)^2}\} ,
\end{equation}
where $\lambda$ is the string tension. In the non-relativistic
limit, such a $G_l$ corresponds to a potential in space configuration,
which has a form as follows

\begin{equation}
\frac{\lambda}{\alpha}(1-e^{-\alpha r}) .
\end{equation}

Since the `on-shell' assumption Eq.(11), the kernel G(P,q,k) in Eq.(5) is $q_0$
and $k_0$ independent. We further define the three dimensional
(instantaneous) wave function as

\begin{equation}
\phi(P,{\bf q})=\int dq_0 \chi(P,q)\; .
\end{equation}
For convenience, we will adopt the center-of-mass frame in
following treatment. In this frame, performing $dq_0$ integration
on both sides of Eq.(5), we obtain the reduced BS equation

\begin{equation}
\phi({\bf q})=(\frac{\Lambda^+({\bf
p_1})\gamma_0}{2E_2(M-E_1-E_2)}
             + \frac{\Lambda^-({\bf p_1})\gamma_0}{2E_2(M+E_1+E_2)})
\int\frac{d^3k}{(2\pi)^3}(-i)G(P,q,k)\phi(P,{\bf k}) ,
\end{equation}
where $\Lambda^\pm({\bf p_1})$ are project operators defined as

\begin{equation}
\Lambda^\pm({\bf p_1})=\frac{E_1\pm\gamma_0({\bf \gamma}
                       \cdot{\bf p_1}+m_1)}{2E_1} .
\end{equation}
To solve the equation, we should construct the
general form of the wave function with the given quantum
numbers. The possible quantum numbers of ($Q\bar{\tilde{b}}$) bound
states are $(\frac{1}{2})^+$, $(\frac{1}{2})^-$,
$(\frac{3}{2})^+$, $(\frac{3}{2})^+$, etc.

For $(\frac{1}{2})^+$ state, the most general
instantaneous wave function in three dimensions has the form

\begin{equation}
\phi({\bf q})=(\phi_1+\phi_2 {\bf \gamma}\cdot {\bf q})u(P) ,
\end{equation}
where $\phi_1$ and $\phi_2$ are scalar functions of $q^2$, and $u(P)$
is the Dirac spinor which satisfies the Dirac equation

\begin{equation}
(P\!\!\!\! /  -M)u(P)=0.
\end{equation}
However, for convenience of the following treatment, we would like
to rewrite the wave function in another form

\begin{equation}
\phi({\bf q})=(\Lambda^+({\bf q})f^{(\frac{1}{2}+)}_1+\Lambda^ -
({\bf q})f^{(\frac{1}{2}+)}_2)u(P) ,
\end{equation}
where $f_1$ and $f_2$ are also scalar functions of $q^2$, which
are related to $\phi_1$ and $\phi_2$ through

\begin{eqnarray}
\phi_1&=&\frac{E_1+m_1}{2E_1}f_1+\frac{E_1-m_1}{2E_1}f_2,\\
\phi_2&=&\frac{1}{2E_1}(f_2-f_1) .
\end{eqnarray}
For $(\frac{1}{2})^-$ state, the three dimensional wave
function is

\begin{equation}
\phi({\bf q})=(\Lambda^+({\bf q})f^{(\frac{1}{2}-)}_1
            +\Lambda^- ({\bf q})f^{(\frac{1}{2}-)}_2)\gamma_5 u(P) .
\end{equation}
For $(\frac{3}{2})^+$ state, the three dimensional wave
function is

\begin{equation}
\phi({\bf q})=(\Lambda^+({\bf q})f^{(\frac{3}{2}+)}_1
     +\Lambda^- ({\bf q})({\bf q})f^{(\frac{3}{2}+)}_2)
        \gamma_5 {\bf q}\cdot {\bf u}(P) ,
\end{equation}
where ${\bf u}(P)$ is a vector spinor which satisfy the following
equations

\begin{equation}
(P\!\!\!\! / -M){\bf u}(P)=0; ~~~~~~   {\bf \gamma}\cdot{\bf u}(P)=0 .
\end{equation}
For $(\frac{3}{2})^-$ state

\begin{equation}
\phi({\bf q})=(\Lambda^+({\bf q})f^{(\frac{3}{2}-)}_1
  +\Lambda^- ({\bf q})f^{(\frac{3}{2}-)}_2) {\bf q}\cdot {\bf u}(P) .
\end{equation}

Inserting the wave function Eq.(22) into Eq.(18), we obtain two
coupled equations:

\begin{eqnarray}
(M-E_1-E_2)\Lambda^+({\bf q})f^{(\frac{1}{2}+)}_1({\bf q})u&=&
\frac{\Lambda^+({\bf q})\gamma_0}{2E_2} \int
\frac{d^3k}{(2\pi)^3}(-i)G(P,{\bf q},{\bf k})
\nonumber \\
& &
\times \left [ \Lambda^+({\bf k})f^{(\frac{1}{2}+)}_1({\bf k})
+\Lambda^- ({\bf k})f^{(\frac{1}{2}+)}_2({\bf k})\right ] u, \\
(M+E_1+E_2)\Lambda^-({\bf q})f^{(\frac{1}{2}+)}_2({\bf q})u&=&
\frac{\Lambda^-({\bf q})\gamma_0}{2E_2} \int
\frac{d^3k}{(2\pi)^3}(-i)G(P,{\bf q},{\bf k})
\nonumber \\
& &\times \left [\Lambda^+({\bf k})f^{(\frac{1}{2}+)}_1({\bf k})
+\Lambda^- ({\bf k})f^{(\frac{1}{2}+)}_2({\bf k})\right ]u.
\end{eqnarray}

Then from these two equations we may abstract two coupled and independent
equations about the two scalar functions $ f^{(\frac{1}{2}+)}_1$ and
$f^{(\frac{1}{2}+)}_2$:

\begin{eqnarray}
(M-E_1-E_2)f^{(\frac{1}{2}+)}_1 &=& \frac{1}{2E_2}\int\frac{d^3k}
     {2E_1^\prime(2\pi)^3}\left\{ G_s^{(1)} [ {\bf k}\cdot({\bf q}+{\bf k})
+\frac{E_1^\prime+m_1}{E_1+m_1}{\bf q}\cdot({\bf q}+{\bf k}) ]
\right .
                               \nonumber \\
& & \hspace*{-1.8cm} \left . +G_s^{(2)}[(E_1^\prime+m_1)+\frac{\bf
q\cdot k}{E_1+m_1}] -G_l[(E_1^\prime+m_1)-\frac{{\bf q\cdot
k}}{E_1+m_1}] \right\}
f^{(\frac{1}{2}+)}_1  \nonumber \\
& & \hspace*{-1.8cm}
 +\frac{1}{2E_2}\int\frac{d^3k}{2E_1^\prime(2\pi)^3}
\left\{ G_s^{(1)} [ -{\bf k}\cdot({\bf q}+{\bf k})
+\frac{E_1^\prime - m_1}{E_1+m_1}{\bf q}\cdot({\bf q}+{\bf k}) ]
\right .
                 \nonumber \\
& &  \hspace*{-1.8cm} \left . +G_s^{(2)}[(E_1^\prime -
m_1)-\frac{\bf q\cdot k}{E_1+m_1}] -G_l[(E_1^\prime - m_1)
+\frac{{\bf q\cdot k}}{E_1+m_1}]
  \right\} f^{(\frac{1}{2}+)}_2 ,\\
(M+E_1+E_2)f^{(\frac{1}{2}+)}_2 &=& \frac{1}{2E_2}
   \int\frac{d^3k}{2E_1^\prime(2\pi)^3}
\left\{ G_s^{(1)} [ {\bf k}\cdot({\bf q}+{\bf k})
-\frac{E_1^\prime+m_1}{E_1-m_1}{\bf q}\cdot({\bf q}+{\bf k}) ]
\right .
                     \nonumber \\
& & \hspace*{-1.8cm} \left . +G_s^{(2)}[(E_1^\prime+m_1)
-\frac{\bf q\cdot k}{E_1-m_1}] -G_l[(E_1^\prime+m_1)+\frac{{\bf
q\cdot k}}{E_1-m_1}] \right\}
f^{(\frac{1}{2}+)}_1  \nonumber \\
& & \hspace*{-1.8cm}
+\frac{1}{2E_2}\int\frac{d^3k}{2E_1^\prime(2\pi)^3} \left\{
G_s^{(1)} [ -{\bf k}\cdot({\bf q}+{\bf k}) -\frac{E_1^\prime -
m_1}{E_1-m_1}{\bf q}\cdot({\bf q}+{\bf k}) ] \right .
             \nonumber \\
& & \hspace*{-1.8cm} \left . +G_s^{(2)}[(E_1^\prime -
m_1)+\frac{\bf q\cdot k}{E_1-m_1}] -G_l[(E_1^\prime - m_1)
-\frac{{\bf q\cdot k}}{E_1-m_1}] \right\} f^{(\frac{1}{2}+)}_2 .
\end{eqnarray}

In a similar way, we can obtain coupled equations for the bound
state with the other quantum numbers, which are shown in the appendix.

\subsection{The bound states of a pair of sbottoms}

The BS wave function for a
$\tilde{b}\bar{\tilde{b}}$ bound state is defined as
\begin{equation}
\chi(P,x_1,x_2)=\langle 0 | T(\tilde{b}(x_1)\bar{\tilde{b}}(x_2))|
P \rangle,
\end{equation}
where $| P \rangle$ is now a bound state of a pair of scalar
bottom quarks. The wave function in momentum space can be defined
similarly as in Eq.(4), then in momentum space the equation
for such a bound state is written as

\begin{equation}
\chi(P,q)=\frac{1}{p^2_1-m^2+i\epsilon}~
\frac{1}{p^2_2-m^2+i\epsilon} \int\frac{d^4
k}{(2\pi)^4}\bar{G}(P,q,k)\chi(P,k),
\end{equation}
where $p_1$ and $p_2$ are momentums of $\tilde{b}$ and
$\bar{\tilde{b}}$ with mass $m$. In the case of the
$\tilde{b}\bar{\tilde{b}}$ bound state, $\lambda_1, \lambda_2$
as in Eqs.(3,6,10) now have $\lambda_1=\lambda_2=\frac{1}{2}$.
The kernel $\bar{G}$ in the above equation is also assumed to
be a combination of two parts
\begin{equation}
\bar{G}=\bar{G}_s+\bar{G}_l .
\end{equation}
The short distance part can be obtained by calculating
one-gluon-exchanged Feynman diagram shown in Fig.~1(b).
\begin{equation}
\bar{G}_s(P,q,k)= i\frac{4}{3}(4\pi \alpha_s)\frac{(p_1+p_3)
\cdot(p_2+p_4)}{(p_2-p_4)^2 - \alpha^2}  .
\end{equation}
The zero components of momentums are again
fixed at their on-shell values. The long distance part of the
kernel is chosen as
\begin{equation}
\bar{G}_l(P,q,k)=
i(2\pi)^3(4m^2_2)\{\frac{\lambda}{\alpha}\delta^3({\bf q} -\bf{k})
-  \frac{\lambda}{\pi^2}\frac{1}{(({\bf q}-{\bf
k})^2+\alpha^2)^2}\} .
\end{equation}
The three dimensional equal time wave function is also defined as
in Eq.(17), and similarly we obtain the reduced instantaneous equation for
the bound state of a pair of scalar bottoms,
\begin{equation}
(M-2E)\phi({\bf q})=\frac{1}{E(M+2E)}
\int\frac{d^3k}{(2\pi)^3}(-i)G(P,q,k)\phi(P,{\bf k}) ,
\end{equation}
where $E=\sqrt{m^2+q^2}$.

Now let us construct the `general form' of the instantaneous wave
function of $\tilde{b}\bar{\tilde{b}}$ bond state. The
possible quantum numbers are $0^{++}$, $1^{--}$, $2^{++}$, etc.
For the state with quantum number $0^{++}$, the three dimensional
wave function is a scalar function
\begin{equation}
\phi({\bf q})=f_0({\bf q}) .
\end{equation}
For the $1^{--}$ state

\begin{equation}
\phi({\bf q})=f_1({\bf q}){\bf q}\cdot{\bf e} ,
\end{equation}
where ${\bf e}$ is the polarization vector of the bound state, and
for $2^{++}$ state the wave function reads

\begin{equation}
\phi({\bf q})=f_2({\bf q})q_iq_j \eta_{ij} ,
\end{equation}
where $\eta_{ij}$ is the polarization tensor of the bound state.
Substituting the wave function Eq.(39) into Eq.(38), we obtain the
equation for $0^{++}$ state. The equations for the states $1^{--}$,
$2^{++}$ and $3^{--}$ may be obtained obtained similarly. These
equations are given explicitly in the appendix.

\section{The numerical results and discussions}

\label{sec3}

Since the integration kernel (potential) is too
complicated for analytically solving, we solve the instantaneous
equations obtained above numerically.

We adopt the method to achieve the low-lying eigenvalues and states
of the equations, first by making the integral equations into discrete ones,
namely to turn the equations into algebra equations, then to make the matrix
for the discrete algebra ones being diagonal.

Besides the masses of quark and scalar quark, there are four
parameters in our investigation: $a$ and $\Lambda_{QCD}$
appearing in the running strong coupling constant, the constant
$\alpha$ and the string tension $\lambda$. These parameters are
fixed by fitting the spectrum of ($Q\bar Q$) bound states in the same
framework of BS equation. Namely to test the reliability of our
numerical method and to choose suitable
parameters appearing in the equations, we calculate the spectrum of
the ordinary heavy quarkonia $(Q\bar{Q})$ and request the calculated values
for the heavy quarkonia are in good agreement with their corresponding
experimental ones. The values for the parameters are fixed as the below values:

$\Lambda_{QCD}=0.162 {\rm GeV}, a=2.713, \lambda=0.23(GeV)^2,
\alpha=0.06 {\rm GeV}, m_b=4.83 {\rm GeV}, m_c=1.55 {\rm GeV}$.

To see the fit of these parameters we have adopted, the calculated spectrum
and the experimental one for heavy quarkonia are put in Table 1. From Table 1
one may conclude how well the chosen parameters in fitting heavy quarkonium
data.

In our investigation, the sbottom mass, $m_{\tilde{b}}$, is
unknown, so one should treat it free. In the present paper, we
constrain ourselves mainly to interest light sbottom with a mass
close to $m_b(\sim 5GeV)$, partly because of the ALEPH
indication\cite{s7} and partly because of interesting scenario to explain
the excess of $b\bar{b}$ pair production in hadron collisions than theoretical
prediction by a factor two. However, in order to
have a full knowledge about the spectrum, we also give the results
for a heavier sbottom. The numerical results about
($Q\bar{\tilde{b}}),(Q=b,c)$ states are given in Table 2 and Table 3.
In Table 4 we show  the results about ($\tilde{b}\bar{\tilde{b}}$).

Because the sbottom is a spin zero particle, the spectrum of the
corresponding bound states is simpler than that of ($Q\bar{Q}$)
states. We can see from these tables that when $m_{\tilde{b}}$ is
close to the mass of bottom quark the masses of the lowest state
of ($b\bar{b}$), ($b\tilde{\bar{b}}$) and ($\tilde{b}\bar{\tilde{b}}$)
are very close to each other. This is reasonable because the
strong interaction for these states is similar. We also see that
the spin-orbit interaction for ($Q\bar{\tilde{b}}$) system is
rather weak: the mass difference between $(\frac{1}{2})^-$ and
$(\frac{3}{2})^-$ is quite small. The reason is that the
quark and scalar quark, being considered here, both are heavy.

The parameters are obtained by fitting the spectrum
for heavy quarkonia ($Q\bar{Q}$). Of the parameters,
we find that the masses of the bound states are not sensitive to the
parameters $a$, $\Lambda_{QCD}$ and $\lambda$. As examples, We
show  $\Lambda_{QCD}$ and $\lambda$ dependence of the masses of
the lowest ($Q\bar{\tilde{b}}$) and ($\tilde{b}\bar{\tilde{b}}$)
bound states in Table 5  and Table 6.

It should be pointed out that the study of the production and
decay properties of such bound states beyond the scope of this
paper and we will be presented elsewhere soon\cite{future}.

\section*{Acknowledgments}

One of the authors (J. Y. Cui) acknowledges the hospitality from
Institute of Theoretical Physics, Chinese Academy of Sciences,
when he visited there. This work was supported in part by Nature Science
Foundation of China (NSFC), and in part by a grant of Chinese Academy of
Science for Outstanding Young Scholars.

\section*{Appendix}

\subsection{The equations of ($Q\bar{\tilde{b}}$) states}

For $(\frac{1}{2})^-$ state the coupled equations are

\begin{eqnarray}
(M-E_1-E_2)f^{(\frac{1}{2}-)}_1 &=&
\frac{1}{2E_2}\int\frac{d^3k}{2E_1^\prime(2\pi)^3} \left\{
G_s^{(1)} [ {\bf k}\cdot({\bf q}+{\bf k})
+\frac{E_1^\prime-m_1}{E_1-m_1}{\bf q}\cdot({\bf q}+{\bf k}) ]
\right . \nonumber\\
& & \hspace*{-2.5cm} \left . +G_s^{(2)}[(E_1^\prime-m_1)+\frac{\bf
q\cdot k}{E_1-m_1}] +G_l[(E_1^\prime-m_1)-\frac{{\bf q\cdot
k}}{E_1-m_1}] \right\}
f^{(\frac{1}{2}-)}_1 \nonumber\\
& & \hspace*{-2.5cm}
 +\frac{1}{2E_2}\int\frac{d^3k}{2E_1^\prime(2\pi)^3}
\left\{ G_s^{(1)} [ -{\bf k}\cdot({\bf q}+{\bf k})
+\frac{E_1^\prime + m_1}{E_1-m_1}{\bf q}\cdot({\bf q}+{\bf k}) ]
\right .
                \nonumber\\
& &  \hspace*{-2.5cm} \left . +G_s^{(2)}[(E_1^\prime +
m_1)-\frac{\bf q\cdot k}{E_1-m_1}] +G_l[(E_1^\prime + m_1)
+\frac{{\bf q\cdot k}}{E_1-m_1}] \right\}
f^{(\frac{1}{2}-)}_2, \\
(M+E_1+E_2)f^{(\frac{1}{2}-)}_2 &=&
 \frac{1}{2E_2}\int\frac{d^3k}{2E_1^\prime(2\pi)^3}
\left\{ G_s^{(1)} [ {\bf k}\cdot({\bf q}+{\bf k})
-\frac{E_1^\prime-m_1}{E_1+m_1}{\bf q}\cdot({\bf q}+{\bf k}) ]
\right .
                    \nonumber\\
& & \hspace*{-2.5cm} \left . +G_s^{(2)}[(E_1^\prime-m_1)
-\frac{\bf q\cdot k}{E_1+m_1}] +G_l[(E_1^\prime-m_1)+\frac{{\bf
q\cdot k}}{E_1+m_1}] \right\}
f^{(\frac{1}{2}-)}_1 \nonumber\\
& &  \hspace*{-2.5cm}
+\frac{1}{2E_2}\int\frac{d^3k}{2E_1^\prime(2\pi)^3} \left\{
G_s^{(1)} [ -{\bf k}\cdot({\bf q}+{\bf k}) -\frac{E_1^\prime +
m_1}{E_1+m_1}{\bf q}\cdot({\bf q}+{\bf k}) ] \right .
                    \nonumber\\
& &  \hspace*{-2.5cm} \left . +G_s^{(2)}[(E_1^\prime +
m_1)+\frac{\bf q\cdot k}{E_1+m_1}] +G_l[(E_1^\prime + m_1)
-\frac{{\bf q\cdot k}}{E_1+m_1}] \right\} f^{(\frac{1}{2}-)}_2 .
\end{eqnarray}
For $(\frac{3}{2})^-$ state, the coupled equations are (in the
following $\theta$ represents the angle between the vectors ${\bf
q}$ and ${\bf k}$):
\begin{eqnarray}
(M-E_1-E_2)f^{(\frac{3}{2}-)}_1 &=&
\frac{1}{2E_2}\int\frac{d^3k}{2E_1^\prime(2\pi)^3} \left\{
G_s^{(1)} [ {\bf k}\cdot({\bf q}+{\bf k})
+\frac{E_1^\prime+m_1}{E_1+m_1}{\bf q}\cdot({\bf q}+{\bf k}) ]
\right . \nonumber\\
& &  \hspace*{-2.5cm} \left .
+G_s^{(2)}[(E_1^\prime+m_1)+\frac{\bf q\cdot k}{E_1+m_1}]
-G_l[(E_1^\prime+m_1)-\frac{{\bf q\cdot k}}{E_1+m_1}] \right\}
\cos\theta  f^{(\frac{3}{2}-)}_1 \nonumber\\
& &  \hspace*{-2.5cm}
+\frac{1}{2E_2}\int\frac{d^3k}{2E_1^\prime(2\pi)^3} \left\{
G_s^{(1)} [ -{\bf k}\cdot({\bf q}+{\bf k}) +\frac{E_1^\prime -
m_1}{E_1+m_1}{\bf q}\cdot({\bf q}+{\bf k}) ] \right .\nonumber\\
& & \hspace*{-2.5cm}
 \left . +G_s^{(2)}[(E_1^\prime - m_1)-\frac{\bf q\cdot k}{E_1+m_1}]
-G_l[(E_1^\prime - m_1) +\frac{{\bf q\cdot k}}{E_1+m_1}] \right\}
\cos\theta  f^{(\frac{3}{2}-)}_2 ,
\end{eqnarray}
\begin{eqnarray}
(M+E_1+E_2)f^{(\frac{3}{2}-)}_2 &=&
\frac{1}{2E_2}\int\frac{d^3k}{2E_1^\prime(2\pi)^3} \left\{
G_s^{(1)} [ {\bf k}\cdot({\bf q}+{\bf k})
-\frac{E_1^\prime+m_1}{E_1-m_1}{\bf q}\cdot({\bf q}+{\bf k}) ]
\right .
                \nonumber\\
& &  \hspace*{-2.5cm} \left . +G_s^{(2)}[(E_1^\prime+m_1)
-\frac{\bf q\cdot k}{E_1-m_1}] -G_l[(E_1^\prime+m_1)+\frac{{\bf
q\cdot k}}{E_1-m_1}] \right\}
\cos\theta f^{(\frac{3}{2}-)}_1 \nonumber\\
& &  \hspace*{-2.5cm}
+\frac{1}{2E_2}\int\frac{d^3k}{2E_1^\prime(2\pi)^3} \left\{
G_s^{(1)} [ -{\bf k}\cdot({\bf q}+{\bf k}) -\frac{E_1^\prime -
m_1}{E_1-m_1}{\bf q}\cdot({\bf q}+{\bf k}) ] \right .\nonumber\\
& &  \hspace*{-2.5cm} \left . +G_s^{(2)}[(E_1^\prime -
m_1)+\frac{\bf q\cdot k}{E_1-m_1}] -G_l[(E_1^\prime - m_1)
-\frac{{\bf q\cdot k}}{E_1-m_1}] \right\} \cos\theta
f^{(\frac{3}{2}-)}_2 .
\end{eqnarray}
For $(\frac{3}{2})^+$ state the coupled equations are
\begin{eqnarray}
(M-E_1-E_2)f^{(\frac{3}{2}+)}_1 &= &
\frac{1}{2E_2}\int\frac{d^3k}{2E_1^\prime(2\pi)^3} \left\{
G_s^{(1)} [ {\bf k}\cdot({\bf q}+{\bf k})
+\frac{E_1^\prime-m_1}{E_1-m_1}{\bf q}\cdot({\bf q}+{\bf k}) ]
\right .\nonumber\\
& &  \hspace*{-2.5cm} \left .
+G_s^{(2)}[(E_1^\prime-m_1)+\frac{\bf q\cdot k}{E_1-m_1}]
+G_l[(E_1^\prime-m_1)-\frac{{\bf q\cdot k}}{E_1-m_1}] \right\}
\cos\theta f^{(\frac{3}{2}+)}_1                 \nonumber\\
& &  \hspace*{-2.5cm}
+\frac{1}{2E_2}\int\frac{d^3k}{2E_1^\prime(2\pi)^3} \left\{
G_s^{(1)} [ -{\bf k}\cdot({\bf q}+{\bf k}) +\frac{E_1^\prime +
m_1}{E_1-m_1}{\bf q}\cdot({\bf q}+{\bf k}) ] \right .
                \nonumber\\
& &  \hspace*{-2.5cm} \left . +G_s^{(2)}[(E_1^\prime +
m_1)-\frac{\bf q\cdot k}{E_1-m_1}] +G_l[(E_1^\prime + m_1)
+\frac{{\bf q\cdot k}}{E_1-m_1}] \right\}
\cos\theta  f^{(\frac{3}{2}+)}_2 ,\\
(M+E_1+E_2)f^{(\frac{3}{2}+)}_2 & =&
\frac{1}{2E_2}\int\frac{d^3k}{2E_1^\prime(2\pi)^3} \left\{
G_s^{(1)} [ {\bf k}\cdot({\bf q}+{\bf k})
-\frac{E_1^\prime-m_1}{E_1+m_1}{\bf q}\cdot({\bf q}+{\bf k}) ]
\right . \nonumber\\
& &  \hspace*{-2.5cm} \left . +G_s^{(2)}[(E_1^\prime-m_1)
-\frac{\bf q\cdot k}{E_1+m_1}] +G_l[(E_1^\prime-m_1)+\frac{{\bf
q\cdot k}}{E_1+m_1}] \right\}
\cos\theta f^{(\frac{3}{2}+)}_1                 \nonumber\\
& &  \hspace*{-2.5cm}
 +\frac{1}{2E_2}\int\frac{d^3k}{2E_1^\prime(2\pi)^3}
\left\{ G_s^{(1)} [ -{\bf k}\cdot({\bf q}+{\bf k})
-\frac{E_1^\prime + m_1}{E_1+m_1}{\bf q}\cdot({\bf q}+{\bf k}) ]
\right .
                \nonumber\\
& &  \hspace*{-2.5cm} \left . +G_s^{(2)}[(E_1^\prime +
m_1)+\frac{\bf q\cdot k}{E_1+m_1}] +G_l[(E_1^\prime + m_1)
-\frac{{\bf q\cdot k}}{E_1+m_1}] \right\} \cos\theta
f^{(\frac{3}{2}+)}_2
\end{eqnarray}

\subsection{The equations of $(\tilde{b}\bar{\tilde{b}})$ states}
For $0^{++}$ state
\begin{equation}
(M-2E)f_0=\frac{1}{E(M+2E)}
\int\frac{d^3k}{(2\pi)^3}(-i)\bar{G}(P,q,k)f_0 .
\end{equation}
For $1^{--}$ state
\begin{equation}
(M-2E)f_1=\frac{1}{E(M+2E)}
\int\frac{d^3k}{(2\pi)^3}(-i)\bar{G}(P,q,k)\cos\theta f_1 .
\end{equation}
For $2^{++}$ state
\begin{equation}
(M-2E)f_2=\frac{1}{E(M+2E)}
\int\frac{d^3k}{(2\pi)^3}(-i)\bar{G}(P,q,k)\frac{3\cos^2\theta
-1}{2}f_2 .
\end{equation}
For $3^{--}$ state
\begin{equation}
(M-2E)f_3=\frac{1}{E(M+2E)}
\int\frac{d^3k}{(2\pi)^3}(-i)\bar{G}(P,q,k)\frac{5\cos^3\theta -
3\cos\theta }{2}f_3 .
\end{equation}

\newpage

\begin{center}
\begin{table}
\caption{The calculated mass spectrum for heavy quarkonia (in
GeV).}
\begin{tabular}{lcc}
    & Calculated & Observed[14] \\ \hline
$1~^1S_0(c\bar c)$  & 2.960 &  2.9788$\pm$ 0.0019\\
$1~^3S_1(c\bar c)$ & 3.100 & 3.09688$\pm$  0.00004 \\
$2~^1S_0 (c\bar c)$ & 3.616 &       \\
$2~^3S_1 (c\bar c)$ & 3.666 & 3.6800$\pm$ 0.00010      \\  \hline
$1~^1S_0 (b\bar b)$ & 9.421 &       \\
$1~^3S_1 (b\bar b)$ & 9.463 &9.46037$\pm$ 0.00021       \\
$2~^1S_0 (b\bar b)$ &9.980 &       \\
$2~^3S_1 (b\bar b)$ & 9.996 &  10.02330$\pm$ 0.00031      \\
$3~^1S_0 (b\bar b)$ &10.331 &       \\
$3~^3S_1 (b\bar b)$ & 10.340 &  10.3553$\pm$0.0005  \\
$4~^1S_0 (b\bar b)$ &10.601 &       \\
$4~^3S_1 (b\bar b)$ & 10.609 &  10.5800$\pm$0.0035 \\ 
\end{tabular}
\end{table}

\begin{table}
\caption{The mass spectrum (in GeV) for $(b\bar{\tilde{b}})$ with
various $m_{\tilde{b}}$ (in GeV).}
\begin{tabular}{lccccc}
 & & n$(\frac{1}{2})^+$ &n$(\frac{1}{2})^-$& n$(\frac{3}{2})^-$ & n$(\frac{3}{2})^+$ \\  \hline
                     & n=1 &  7.659 &  8.054   &  8.067  &  8.121    \\
$m_{\tilde{b}}$=3.0 & n=2 &  8.218 & 8.444    &  8.453   &  8.424    \\
                    & n=3 &  8.575 & 8.741   &  8.747    &  8.800   \\    \hline
                    & n=1 & 8.145 &  8.538   &  8.551  &  8.604    \\
$m_{\tilde{b}}$=3.5 & n=2 &  8.700 & 8.925    &  8.932   & 9.003    \\
                    & n=3 &  9.054 & 9.219   &  9.224    & 9.278   \\    \hline
                    & n=1 & 8.633 &  9.024   &  9.036  &  9.089    \\
$m_{\tilde{b}}$=4.0 & n=2 &  9.184 & 9.407    &  9.415   & 9.485    \\
                    & n=3 &  9.535 & 9.699   &  9.704    & 9.758   \\    \hline
                    & n=1 & 9.122  & 9.512   & 9.523   &  9.576    \\
$m_{\tilde{b}}$=4.5 & n=2 & 9.670  & 9.892    &  9.899   & 9.969    \\
                    & n=3 & 10.019  & 10.182   & 10.187     & 10.240   \\    \hline
                     & n=1 & 9.613   & 10.001   & 10.012   &  10.064    \\
$m_{\tilde{b}}$=5.0 & n=2 & 10.158  & 10.379    &10.376   & 10.455    \\
                    & n=3 & 10.505  & 10.667   & 10.672   &10.725   \\    \hline
                    & n=1 & 10.104  & 10.491   & 10.502    & 10.554    \\
$m_{\tilde{b}}$=5.5 & n=2 & 10.648  & 10.867    & 10.874    & 10.943    \\
                    & n=3 & 10.992  & 11.154   & 11.158    & 11.211   \\    \hline
                    & n=1 & 10.596  & 10.983    & 10.993   & 11.045    \\
$m_{\tilde{b}}$=6.0 & n=2 & 11.138  & 11.357    &  11.363   & 11.432    \\
                    & n=3 & 11.481   & 11.642   &  11.646    & 11.699   \\    \hline
                    & n=1 & 12.573   &  12.956   & 12.965   & 13.016     \\
$m_{\tilde{b}}$=8.0 & n=2 & 13.109   &  13.325   & 13.330   & 13.400    \\
                    & n=3 &  13.447   &  13.605  &  13.608    & 13.661   \\    \hline
                    & n=1 &  14.557   &  14.938   & 14.946   & 14.996     \\
$m_{\tilde{b}}$=10.0 & n=2 & 15.090   & 15.302    & 15.307    & 15.374    \\
                    & n=3  & 15.424   &  15.580  &  15.583    &  15.634  \\    \hline
                    & n=1 &   24.519    & 24.895    &24.902    & 24.950     \\
$m_{\tilde{b}}$=20.0 & n=2 & 25.045   & 25.251    & 25.255    &25.320     \\
                    & n=3  & 25.371   &  25.521  &  25.523    & 25.573   \\    \hline
                    & n=1 &   44.497    & 44.871    & 44.878   & 44.922     \\
$m_{\tilde{b}}$=40.0 & n=2 &  45.020   & 45.222    & 45.230    &45.288     \\
                    & n=3  &  45.339   & 45.487   &  45.496    &45.554    \\    \hline
                    & n=1 &   64.492    & 64.865    & 64.869   & 64.914     \\
$m_{\tilde{b}}$=60.0 & n=2 &  65.015   & 65.218    & 65.219    & 65.378    \\
                    & n=3  & 65.336   & 65.483   &  65.484    & 65.524   \\ 
\end{tabular}
\end{table}

\begin{table}
\caption{The mass spectrum (in GeV) for $(c\bar{\tilde{b}})$ with
various $m_{\tilde{b}}$ (in GeV).}
\begin{tabular}{lccccc} 
~~~&~~~~&\hskip0.2in  n$(\frac{1}{2})^+$\hskip 0.2in~ &\hskip0.2in n$(\frac{1}{2})^-$\hskip 0.2in~ &
\hskip 0.2in n$(\frac{3}{2})^-$\hskip 0.2in~ &\hskip 0.2in n$(\frac{3}{2})^+$\hskip0.2in~  \\  \hline
                    & n=1 & 4.530   & 4.906   & 4.931   & 4.994     \\
$m_{\tilde{b}}$=3.0 & n=2 & 5.091  & 5.317   &5.331    &5.406      \\
                    & n=3 & 5.458  & 5.625   &5.635   &5.691    \\    \hline
                    & n=1 & 5.029  & 5.402    & 5.425   & 5.488     \\
$m_{\tilde{b}}$=3.5 & n=2 & 5.583  & 5.808    &  5.821   & 5.895    \\
                    & n=3 &5.947  & 6.112    & 6.122    &6.177   \\    \hline
                    & n=1 &5.527  & 5.899    & 5.919   & 5.983    \\
$m_{\tilde{b}}$=4.0 & n=2 &6.077  & 6.300   & 6.312   & 6.386    \\
                    & n=3 &6.436 & 6.601   & 6.610   & 6.665   \\    \hline
                    & n=1 &6.025 &  6.395  & 6.414  &  6.478   \\
$m_{\tilde{b}}$=4.5 & n=2 &6.570   &6.793  & 6.804   & 6.877   \\
                    & n=3  &6.927  &7.091  & 7.099   & 7.155  \\    \hline
                    & n=1 & 6.523 & 6.892    &6.910    &6.974      \\
$m_{\tilde{b}}$=5.0 & n=2 &7.065   &7.286    &7.297    &7.370      \\
                    & n=3  &7.419   &7.582   &7.590    &7.645    \\ \hline
                    & n=1 &7.021  & 7.389    & 7.406   & 7.469     \\
$m_{\tilde{b}}$=5.5 & n=2 &7.561   &7.781    & 7.791    &7.864     \\
                    & n=3 &7.912   &8.075    &8.082     &8.137   \\    \hline
                    & n=1 &7.519  &7.887     & 7.902   & 7.966    \\
$m_{\tilde{b}}$=6.0 & n=2 &8.056  &8.276    & 8.285   & 8.358    \\
                    & n=3  &8.406   &8.568    & 8.575    &8.629    \\    \hline
                    & n=1 &9.513   & 9.878       & 9.891   & 9.955    \\
$m_{\tilde{b}}$=8.0 & n=2 &10.044  & 10.260    & 10.268    &10.340    \\
                    & n=3  &10.387  & 10.547   & 10.552     &10.606   \\    \hline
                     & n=1 &11.509  & 11.872    &11.884    & 11.947     \\
$m_{\tilde{b}}$=10.0 & n=2 &12.035   & 12.250    &12.257     &12.328      \\
                     & n=3  & 12.375  & 12.532   & 12.537     &12.591    \\    \hline
                     & n=1 &21.500  & 21.859    &21.867    & 21.929     \\
$m_{\tilde{b}}$=20.0 & n=2 & 22.016  & 22.226   & 22.231    & 22.301    \\
                     & n=3  &22.346   & 22.500   & 22.503    & 22.555  \\    \hline
                     & n=1 &41.494  & 41.851    &41.857    & 41.919    \\
$m_{\tilde{b}}$=40.0 & n=2 &42.005  & 42.213   & 42.216   &42.286     \\
                     & n=3  & 42.330  & 42.481   & 42.483  & 42.535   \\    \hline
                     & n=1 &61.490 & 61.848   & 61.854  & 61.914    \\
$m_{\tilde{b}}$=60.0 & n=2 &62.003   & 62.208    & 62.211    &26.280    \\
                     & n=3 &62.325   & 62.474   & 62.476     &62.528   \\ 
\end{tabular}
\end{table}

\begin{table}
\caption{The mass spectrum (in GeV) for
$(\tilde{b}\bar{\tilde{b}})$ with various $m_{\tilde{b}}$ (in
GeV).}
\begin{tabular}{lccccc} 
~~~&~~~~&\hskip0.2in  n$0^{++}$\hskip 0.2in~ &\hskip0.2in n$1^{--}$\hskip 0.2in~ &
\hskip 0.2in n$2^{++}$\hskip 0.2in~ &\hskip 0.2in n$3^{--}$\hskip0.2in~  \\  \hline
                    & n=1 & 5.969  &6.313    &6.553    & 6.742     \\
$m_{\tilde{b}}$=3.0 & n=2 &6.454   &6.668    &6.844    & 6.993     \\
                    & n=3 &6.779   &6.939    &7.079     &7.201    \\    \hline
                    & n=1 &6.930   &7.280    &7.520      &7.709      \\
$m_{\tilde{b}}$=3.5 & n=2 &7.417   &7.634    &7.810     & 7.959    \\
                    & n=3 &7.742   &7.904    &8.043     & 8.167  \\    \hline
                    & n=1 &7.895   &8.250    &8.489     & 8.678    \\
$m_{\tilde{b}}$=4.0 & n=2 &8.383   &8.601     &8.777     &8.926     \\
                    & n=3 &8.707   &8.869    & 9.010    & 9.133  \\    \hline
                    & n=1 &8.863   &9.221    & 9.461    & 9.649    \\
$m_{\tilde{b}}$=4.5 & n=2 &9.353   &9.570    & 9.746   &  9.895    \\
                    & n=3 &9.675   &9.837    & 9.978    &  10.102  \\    \hline
                    & n=1 &9.833   &10.195    &10.434      &10.621      \\
$m_{\tilde{b}}$=5.0 & n=2 &10.324   &10.542    &10.717     &10.865     \\
                    & n=3 &10.645   &10.807    &10.947     &11.071   \\    \hline
                    & n=1 &10.805   &11.170    &11.409     &11.595     \\
$m_{\tilde{b}}$=5.5 & n=2 &11.298   &11.515     &11.690     &11.837     \\
                    & n=3 &11.617   &11.778    & 11.918    & 12.042  \\    \hline
                    & n=1 &11.779   &12.147    & 12.385    & 12.571    \\
$m_{\tilde{b}}$=6.0 & n=2 &12.273   &12.490    & 12.664   &  12.811    \\
                    & n=3 &12.590   &12.752    & 12.892    & 13.014   \\    \hline
                    & n=1 &15.689   &16.069    & 16.305     & 16.496     \\
$m_{\tilde{b}}$=8.0 & n=2 &16.189   &16.405    & 16.576    &  16.720   \\
                    & n=3 &16.500   &16.660    & 16.798    &  16.918 \\    \hline
                    & n=1 &19.615   &20.005    & 20.240    &  20.420   \\
$m_{\tilde{b}}$=10.0 & n=2 &20.122   &20.338     &20.506     &20.647     \\
                    & n=3 &20.428   &20.586    &20.721     &  20.839 \\    \hline
                    & n=1 &39.374   &39.801    &40.037     &  40.208   \\
$m_{\tilde{b}}$=20.0 & n=2 &39.918   &40.127    &40.287    &  40.420    \\
                    & n=3 &40.213   &40.359    &40.484     &  40.594  \\    \hline
                    & n=1 &79.146   &79.602    &79.842      & 80.010     \\
$m_{\tilde{b}}$=40.0 & n=2 &79.745   &79.941    &80.009     & 80.216    \\
                    & n=3 &80.032   & 80.165   & 80.280    &  80.380 \\    \hline
                    & n=1 &119.035   &119.500    &119.742     &119.909     \\
$m_{\tilde{b}}$=60.0 & n=2 &119.667   &119.855     &120.000     &120.120     \\
                    & n=3 &119.950   &120.078    & 120.186    &  120.283 \\ 
\end{tabular}
\end{table}

\begin{table}
\caption{The mass spectrum dependence of $\Lambda_{QCD}$.}
\begin{tabular}{lcc} 
\hskip4mm $GeV$ &\hskip0.8in $(b\tilde{b})(\frac{1}{2})^+ (GeV)$ \hskip0.6in~ & \hskip0.6in
$(\tilde{b}\bar{\tilde{b}})(0^{++})(GeV)$\hskip0.6in~\\  \hline
$\Lambda_{QCD}=0.11 $ &  9.745  &9.943 \\
$\Lambda_{QCD}=0.13 $ &  9.692  &9.898\\
$\Lambda_{QCD}=0.15 $ &  9.641  &9.85\\
$\Lambda_{QCD}=0.17 $ &  9.594  &9.817\\
$\Lambda_{QCD}=0.19 $ &  9.577  &9.780\\
$\Lambda_{QCD}=0.21 $ &  9.504  &9.744\\
$\Lambda_{QCD}=0.23 $ &  9.461  &9.711\\
$\Lambda_{QCD}=0.25 $ &  9.420  &9.679 \\
\end{tabular}
\end{table}

\begin{table}
\caption{The mass spectrum dependence of $\lambda(GeV^2)$.}
\begin{tabular}{lcc} 
\hskip3mm $GEV^2$&\hskip0.9in $(b\tilde{b})(\frac{1}{2})^+(GeV)$ \hskip0.6in~ & \hskip0.6in
$(\tilde{b}\bar{\tilde{b}})(0^{++})(GeV) $\hskip0.6in~\\  \hline
$\lambda=0.15 $ &  9.541  &9.764 \\
$\lambda=0.16 $ &  9.550  &9.773\\
$\lambda=0.17 $ &  9.559  &9.782\\
$\lambda=0.18 $ &  9.568  &9.790\\
$\lambda=0.19 $ &  9.578  &9.799\\
$\lambda=0.20 $ &  9.586  &9.807\\
$\lambda=0.21 $ &  9.595  &9.816\\
$\lambda=0.22 $ &  9.604  &9.824 \\
$\lambda=0.23 $ &  9.612  &9.832\\
$\lambda=0.24 $ &  9.621  &9.840\\
$\lambda=0.25 $ &  9.630  &9.848 \\
\end{tabular}
\end{table}
\end{center}
\end{document}